\documentclass[letterpaper,twocolumn,english,aps,prl,showpacs,superscriptaddress, floatfix]{revtex4-1}
\usepackage[T1]{fontenc}
\usepackage[latin9]{inputenc}
\usepackage{color}
\usepackage{babel}
\usepackage{amsmath}
\usepackage{amssymb}
\usepackage{graphicx}
\usepackage[unicode=true,
 bookmarks=false,
 breaklinks=true,pdfborder={0 0 0},backref=false,colorlinks=true]
 {hyperref}
\hypersetup{pdftitle={Beam-Target Double Spin Asymmetry A_LT in Charged Pion Production from Deep Inelastic Scattering on a Transversely Polarized He-3 Target at 1.4<Q^2<2.7 GeV^2 },
 pdfauthor={J. Huang, et. al. (The Jefferson Lab Hall A Collaboration)},
 pdfsubject={v22: nucl-ex, hep-ex},
 pdfkeywords={DOI:},
 linkcolor=DarkBlueCite, citecolor=DarkBlueCite, urlcolor = DarkBlueCite}

\makeatletter

\pdfpageheight\paperheight
\pdfpagewidth\paperwidth

\@ifundefined{textcolor}{}
{%
 \definecolor{BLACK}{gray}{0}
 \definecolor{WHITE}{gray}{1}
 \definecolor{RED}{rgb}{1,0,0}
 \definecolor{GREEN}{rgb}{0,1,0}
 \definecolor{BLUE}{rgb}{0,0,1}
 \definecolor{CYAN}{cmyk}{1,0,0,0}
 \definecolor{MAGENTA}{cmyk}{0,1,0,0}
 \definecolor{YELLOW}{cmyk}{0,0,1,0}
 }

\usepackage{dcolumn}
\usepackage{bm}

\usepackage{ifpdf} 
\ifpdf 

 \IfFileExists{lmodern.sty}{\usepackage{lmodern}}{}

\fi 
\definecolor{DarkBlueCite}{rgb}{0.1,0.0,0.5}

\newcommand{\moeller}{M{\o}ller }

\usepackage{dcolumn}

\makeatother

\begin{document}


\author{J.~Huang}\email[Corresponding author: ]{jinhuang@jlab.org}
\affiliation{Massachusetts Institute of Technology, Cambridge, Massachusetts 02139, USA}
\author{K.~Allada}
\affiliation{University of Kentucky, Lexington, Kentucky 40506, USA}
\author{C.~Dutta}
\affiliation{University of Kentucky, Lexington, Kentucky 40506, USA}
\author{J.~Katich}
\affiliation{College of William and Mary, Williamsburg, Virginia 23187, USA}
\author{X.~Qian} 
\affiliation{Duke University, Durham, North Carolina 27708, USA}
\affiliation{Kellogg Radiation Laboratory, California Institute 
  of Technology, Pasadena, California 91125, USA}
\author{Y.~Wang}
\affiliation{University of Illinois at Urbana-Champaign, Urbana, illinois 61801, USA}
\author{Y.~Zhang}
\affiliation{Lanzhou University, Lanzhou 730000, Gansu, People's Republic of China}
%
\author{K.~Aniol}
\affiliation{California State University, Los Angeles, Los Angeles, California 90032, USA}
\author{J.R.M.~Annand}
\affiliation{University of Glasgow, Glasgow G12 8QQ, Scotland, United Kingdom}
\author{T.~Averett}
\affiliation{College of William and Mary, Williamsburg, Virginia 23187, USA}
% ...
\collaboration{The Jefferson Lab Hall A Collaboration}
\noaffiliation

\preprint{\textbf{v22.0}, PRL Final - ArXiv V2}

\title{Beam-Target Double Spin Asymmetry $A_{LT}$ in Charged Pion Production
from Deep Inelastic Scattering on a Transversely Polarized $^{3}$He
Target at $1.4<Q^{2}<2.7\,\textrm{GeV}^{2}$}

\date{\today}
\begin{abstract}
We report the first measurement of the double-spin asymmetry $A_{LT}$
for charged pion electroproduction in semi\nobreakdash-inclusive
deep\nobreakdash-inelastic electron scattering on a transversely
polarized $^{3}$He target. The kinematics focused on the valence
quark region, $0.16<x<0.35$ with $1.4<Q^{2}<2.7\,\textrm{GeV}^{2}$.
The corresponding neutron $A_{LT}$ asymmetries were extracted from
the measured $^{3}$He asymmetries and proton over $^{3}$He cross
section ratios using the effective polarization approximation. These
new data probe the transverse momentum dependent parton distribution
function $g_{1T}^{q}$ and therefore provide access to quark spin-orbit
correlations. Our results indicate a positive azimuthal asymmetry
for $\pi^{-}$ production on $^{3}$He and the neutron, while our
$\pi^{+}$ asymmetries are consistent with zero.
\end{abstract}

\pacs{14.20.Dh, 25.30.Fj, 25.30.Rw, 24.85.+p}

\maketitle
Understanding the spin structure of the nucleon in terms of parton
spin and orbital angular momentum (OAM) remains a fundamental challenge
in contemporary hadronic physics. The transverse momentum dependent
(TMD) parton distribution functions (PDFs)~\citep{Mulders:1995dh,Boer:1997nt}
describe the spin\nobreakdash-correlated three-dimensional momentum
structure of the nucleon's quark constituents. Of the eight leading\nobreakdash-twist
TMD PDFs, five vanish after integration over quark's transverse momentum,
$\boldsymbol{p}_{T}$. Experimental information on these TMD PDFs
is rather scarce. Among them, the transversal helicity $g_{1T}^{q}$
is a T\nobreakdash-even and chiral\nobreakdash-even distribution,
which describes the $\boldsymbol{p}_{T}$\nobreakdash-correlated
longitudinal polarization of quarks in a transversely polarized nucleon~\citep{Mulders:1995dh,Kotzinian:1995cz}.
Because $g_{1T}^{q}$ requires an interference between wave function
components differing by one unit of quark OAM~\citep{Ji:2002xn},
the observation of a nonzero $g_{1T}^{q}$ would provide direct evidence
that quarks carry orbital angular momentum, constraining an important
part of the nucleon spin sum rule~\citep{Jaffe:1989jz}.

In recent years, semi\nobreakdash-inclusive deep\nobreakdash-inelastic
lepton\nobreakdash-nucleon scattering~(SIDIS) and the Drell\nobreakdash-Yan
process have been recognized as clean experimental probes for TMD
PDFs~\citep{Barone:2010zz}. In the SIDIS process, $\ell(l)+N(P)\to\ell(l')+h(P_{h})+X$,
a lepton ($\ell$) scatters from a nucleon ($N$) and is detected
in coincidence with a leading hadron ($h$) with particle four-momenta
denoted by $l$, $P$, $l'$ and $P_{h}$, respectively. All eight
leading\nobreakdash-twist TMD PDFs can be accessed using SIDIS~\citep{Bacchetta:2006tn}.
In particular, the beam\nobreakdash-helicity double\nobreakdash-spin
asymmetry (DSA) $A_{LT}$ in SIDIS reactions on a transversely polarized
nucleon is given at leading twist by 
\begin{align}
A_{LT}\left(\phi_{h},\phi_{S}\right) & \equiv\frac{1}{\left|P_{B}S_{T}\right|}\frac{Y^{+}\left(\phi_{h},\phi_{S}\right)-Y^{-}\left(\phi_{h},\phi_{S}\right)}{Y^{+}\left(\phi_{h},\phi_{S}\right)+Y^{-}\left(\phi_{h},\phi_{S}\right)}\label{eq:ALT.Def}\\
 & \approx A_{LT}^{\cos\left(\phi_{h}-\phi_{S}\right)}\cos\left(\phi_{h}-\phi_{S}\right),\nonumber 
\end{align}
where $\phi_{h}$ and $\phi_{S}$ are the azimuthal angles of the
produced hadron and the target spin as defined in the Trento convention~\citep{Bacchetta:2004jz},
$P_{B}$ is the polarization of the lepton beam, $S_{T}$ is the transverse
polarization of the target, and $Y^{\pm}\left(\phi_{h},\phi_{S}\right)$
is the normalized yield for beam helicity of $\pm1$. The first and
second subscripts to $A$ denote the respective polarization of beam
and target (L, T, and U represent longitudinal, transverse, and unpolarized,
respectively). The partonic interpretation of the SIDIS cross section
at the kinematic region of this experiment is supported by QCD factorization
theory~\citep{Ji:2004wu} and experimental data~\citep{Avakian:2003pk,Asaturyan:2011mq}.
At leading order~(LO), the $A_{LT}^{\cos\left(\phi_{h}-\phi_{S}\right)}$
asymmetry is proportional to the convolution of $g_{1T}^{q}$ and
the unpolarized fragmentation function (FF) $D_{1}$~\citep{Kotzinian:1995cz,Bacchetta:2006tn}.

Significant progress in theory and phenomenology regarding $g_{1T}^{q}$
and the related $A_{LT}^{\cos(\phi_{h}-\phi_{S})}$ asymmetry have
been achieved in recent years. In a light\nobreakdash-cone constituent
quark model~\citep{Boffi:2009sh}, $g_{1T}^{q}$ is explicitly decomposed
into a dominant contribution from the interference of S- and P-waves
and a minor ($<20\%$) contribution from the interference of P\nobreakdash-
and D\nobreakdash- waves in the quark wavefunctions. The $\boldsymbol{p}_{T}^{2}$\nobreakdash-moment
of $g_{1T}^{q}$ can be estimated from the collinear $g_{1}^{q}$
distribution function~\citep{deFlorian:2008mr} using the Wandzura-Wilczek~(WW)\nobreakdash-type
approximation~\citep{Mulders:1995dh,Kotzinian:1995cz}, which neglects
the higher\nobreakdash-twist contributions. In addition, the TMD
PDFs have recently been explored in lattice QCD, using a simplified
definition of the TMD PDFs with straight gauge links~\citep{Hagler:2009mb}.
$g_{1T}^{q}$ was among the first TMD PDFs addressed with this method.
$g_{1T}^{q}$ has also been calculated in quark models as discussed
in Refs.~\citep{Jakob:1997wg,Pasquini:2008ax,Kotzinian:2008fe,Efremov:2009ze,Avakian:2010br,Bacchetta:2010si,Zhu:2011zz,Efremov:2011ye}.
Common features of these models suggest that $g_{1T}^{u}$ is positive
and $g_{1T}^{d}$ is negative. Both reach their maxima in the valence
region at the few-percent level relative to the unpolarized distribution
$f_{1}^{q}$. The simple relation $g_{1T}^{q}=-h_{1L}^{\perp q}$,
where the $h_{1L}^{\perp q}$ TMD PDF leads to the SIDIS $A_{UL}$
asymmetry, has an essentially geometric origin and is supported by
a large number of models~\citep{Lorce:2011zt}. Moreover, recent
lattice QCD calculations indicate that the relation may indeed be
approximately satisfied~\citep{Hagler:2009mb,Musch:2010ka}. In addition,
the QCD parton model suggests approximate TMD relations, which link
$g_{1T}^{q}$ with the quark transversity distribution $h_{1}^{q}$
and the pretzelosity distribution, $h_{1T}^{\perp q}$~\citep{DiSalvo:2006wf}.
$A_{LT}^{\cos\left(\phi_{h}-\phi_{S}\right)}$ has been predicted
for the kinematics and reaction channels of this experiment using
the WW\nobreakdash-type approximations~\citep{Kotzinian:2006dw,Prokudin.DraftPaper},
a light\nobreakdash-cone constituent quark model~\citep{Pasquini:2008ax,Boffi:2009sh},
a diquark spectator model~\citep{Bacchetta:2010si} and a light\nobreakdash-cone
quark\nobreakdash-diquark model~\citep{Zhu:2011zz}.

The COMPASS collaboration previously reported preliminary results
for $A_{LT}^{\cos\left(\phi_{h}-\phi_{S}\right)}$ in positive and
negative charged hadron production using a muon beam scattered from
transversely polarized deuterons~\citep{Parsamyan:2007ju} and protons~\citep{Parsamyan:2010se}.
The kinematics favored the sea quark region. Within the uncertainties,
the preliminary results cannot differentiate between zero and various
model predictions.

In this Letter, we report new results from experiment E06-010 in Jefferson
Lab Hall A, which measured the $A_{LT}$ DSA and the target single
spin asymmetries (target\nobreakdash-SSA)~\citep{Qian:2011py} in
SIDIS reactions on a transversely polarized $^{3}$He target. The
experiment used a longitudinally polarized 5.9~GeV electron beam
with an average current of 12$\,\mu$A. Polarized electrons were excited
from a superlattice GaAs photocathode by a circularly polarized laser~\citep{Sinclair:2007ez}
at the injector of the CEBAF accelerator. The laser polarization,
and therefore the electron beam\nobreakdash-helicity, was flipped
at 30 Hz using a Pockels cell. The average beam polarization was $\left(76.8\pm3.5\right)\%$,
which was measured periodically by \moeller polarimeter. Through
an active feedback system~\citep{Androic:2011rha}, the beam charge
asymmetry between the two helicity states was controlled to less than
150~ppm over a typical 20~min period between target spin\nobreakdash-flips
and less than 10~ppm for the entire experiment. In addition to the
fast helicity flip, roughly half of the data were accumulated with
a half-wave plate inserted in the path of the laser at the source,
providing a passive helicity reversal for an independent cross\nobreakdash-check
of the systematic uncertainty. 

The ground state $^{3}$He wave function is dominated by the S state,
in which the two proton spins cancel and the nuclear spin resides
entirely on the single neutron~\citep{Bissey:2001cw}. Therefore,
a polarized $^{3}$He target is the optimal effective polarized neutron
target. The target used in this measurement is polarized by spin-exchange
optical pumping of a Rb-K mixture~\citep{Babcock:2003zz}. A significant
improvement in target polarization compared to previous experiments
was achieved using spectrally narrowed pumping lasers~\citep{Singh:2009zzk},
which improved the absorption efficiency. The $^{3}$He gas of $~10$~atm
pressure was contained in a 40\nobreakdash-cm\nobreakdash-long glass
vessel, which provided an effective electron-polarized neutron luminosity
of $10^{36}\text{\,}\text{cm}^{-2}\mbox{s}^{-1}$. The beam charge
was divided equally among two target spin orientations transverse
to the beamline, parallel and perpendicular to the central $\vec{l}$-$\vec{l}'$
scattering plane. Within each orientation, the spin direction of the
$^{3}$He was flipped every 20~min through adiabatic fast passage~\citep{Abragam1961.NMR}.
The average in-beam polarization was $(55.4\pm2.8)\%$ and was measured
during each spin flip using nuclear magnetic resonance, which in turn
was calibrated regularly using electron paramagnetic resonance~\citep{Romalis:1998}.

The scattered electron was detected in the BigBite spectrometer, which
consisted of a single dipole magnet for momentum analysis, three multiwire
drift chambers for tracking, a scintillator plane for time\nobreakdash-of\nobreakdash-flight
measurement and a lead\nobreakdash-glass calorimeter divided into
preshower and shower sections for electron identification (ID) and
triggering. Its angular acceptance was about 64~msr for a momentum
range from 0.6 to 2.5~GeV. The left high resolution spectrometer
(HRS)~\citep{Alcorn:2004sb} was used to detect hadrons in coincidence
with the BigBite spectrometer. Its detector package included two drift
chambers for tracking, two scintillator planes for timing and triggering,
a gas Cerenkov detector and a lead\nobreakdash-glass shower detector
for electron ID. In addition, an aerogel \v{C}erenkov detector and
a ring imaging \v{C}erenkov detector were used for hadron ID. The
HRS central momentum was fixed at 2.35~GeV with a momentum acceptance
of $\pm4.5\%$ and an angular acceptance of $\sim$6~msr.

The SIDIS event sample was selected with particle identification and
kinematic cuts, including the four momentum transfer squared $Q^{2}>1$
GeV$^{2}$, the virtual\nobreakdash-photon\nobreakdash-nucleon invariant
mass $W>2.3$ GeV, and the mass of undetected final-state particles
$W^{\prime}>1.6\textrm{ GeV}$. The kinematic coverage was in the
valence quark region for values of the Bjorken scaling variable in
$0.16<x<0.35$ at a scale of $1.4<Q^{2}<2.7\,\textrm{GeV}^{2}$. The
range of measured hadron transverse momentum $P_{h\perp}$ was $0.24$-$0.44$~GeV.
The fraction $z$ of the energy transfer carried by the observed hadron
was confined by the HRS momentum acceptance to a small range about
$z\sim0.5\mbox{-}0.6$. Events were divided into four $x$ bins with
equivalent statistics. At high $x$, the azimuthal acceptance in $\phi_{h}-\phi_{S}$
was close to 2$\pi$, while at lower $x$, roughly half of the $2\pi$
range was covered, including the regions of maximal and minimal sensitivity
to $A_{LT}^{\cos\left(\phi_{h}-\phi_{S}\right)}$ at $\cos\left(\phi_{h}-\phi_{S}\right)\sim\pm1$
and zero, respectively. The central kinematics were presented in Ref.~\citep{Qian:2011py}.

\begin{figure}
\includegraphics[bb=0.17999999999999999in 2.1000000000000001in 10in 8in,clip,width=9cm]{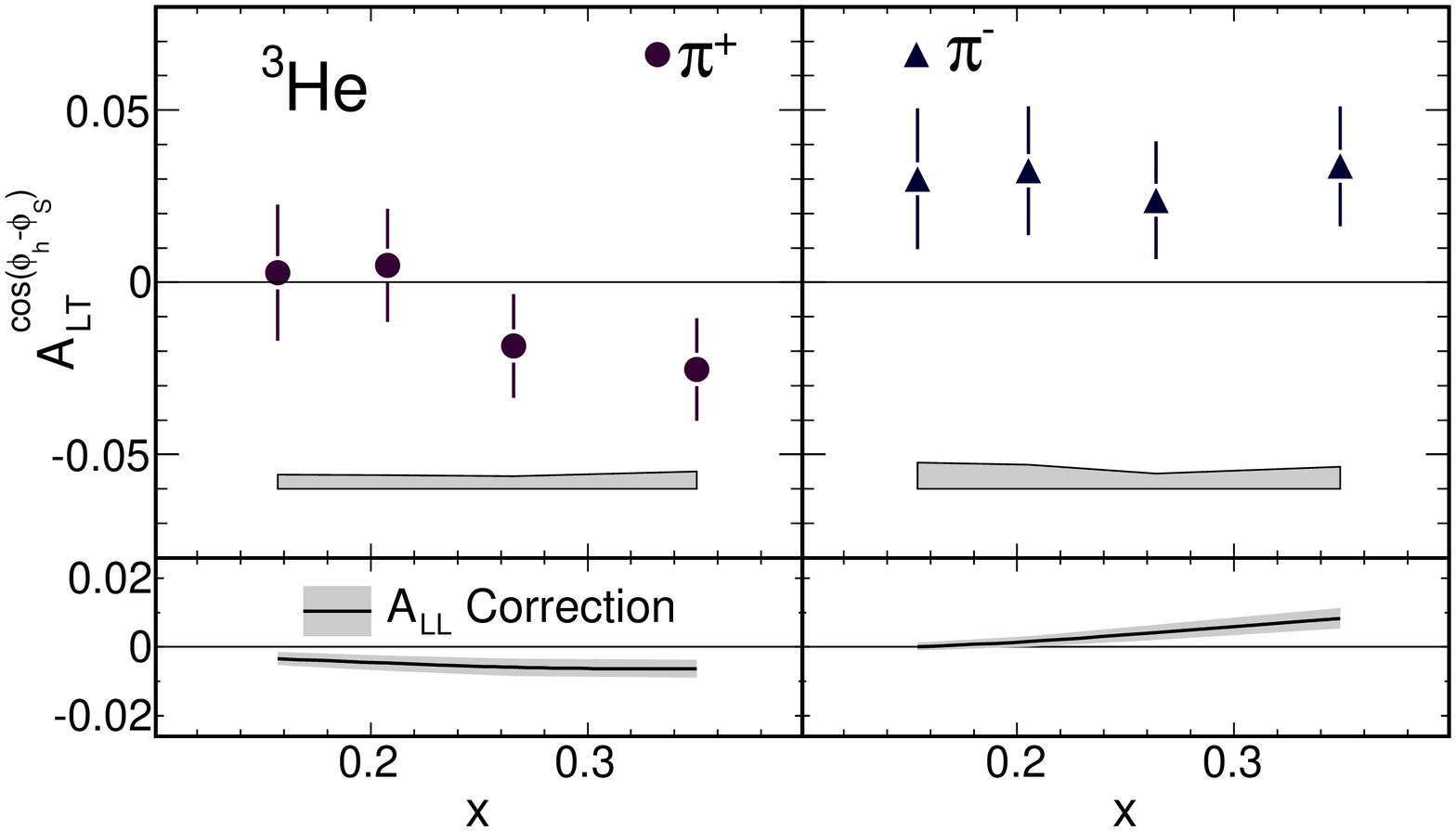}

\caption{\label{fig:He--azimuthal}$^{3}$He $A_{LT}^{\cos\left(\phi_{h}-\phi_{S}\right)}$
azimuthal asymmetry plotted against $x$ for positive (top left) and
negative (top right) charged pions. The $A_{LL}$ correction (see
text) that was applied and its uncertainty are shown in the bottom
panels.}
\end{figure}
The beam-helicity DSA was formed from the measured yields as in Eq.~\eqref{eq:ALT.Def}.
The azimuthal asymmetry in each $x$\nobreakdash-bin was extracted
directly using an azimuthally unbinned maximum likelihood estimator
with corrections for the accumulated beam charge, the data acquisition
live time, and the beam and target polarizations. The result was confirmed
by an independent binning\nobreakdash-and\nobreakdash-fitting procedure~\citep{Qian:2011py}.
The sign of the asymmetry was cross\nobreakdash-checked with that
of the known asymmetry of $^{3}\vec{\mbox{He}}(\vec{e},e')$ elastic
and quasielastic scattering on longitudinally and transversely polarized
targets~\citep{Donnelly:1985ry}. The small amount of unpolarized
$\text{N}_{2}$ used in the target cell to reduce depolarization diluted
the measured $^{3}$He asymmetry, which was corrected for the nitrogen
dilution defined as 
\begin{equation}
f_{\text{N}_{2}}\equiv\frac{N_{\text{N}_{2}}\sigma_{\text{N}_{2}}}{N_{^{3}\text{He}}\sigma_{^{3}\text{He}}+N_{\text{N}_{2}}\sigma_{\text{N}_{2}}},
\end{equation}
where $N$ is the density and $\sigma$ is the unpolarized SIDIS cross
section. The ratio $\sigma_{^{3}\text{He}}/\sigma_{\text{N}_{2}}$
was measured periodically in dedicated runs on targets filled with
known amounts of pure unpolarized $^{3}$He and $\text{N}_{2}$, resulting
$f_{\text{N}_{2}}\sim10\%$. A $5\%\text{-}20\%$ longitudinal component
of the target polarization with respect to the virtual\nobreakdash-photon
direction introduced a small correction to $A_{LT}\left(\phi_{h},\phi_{S}\right)$
from the DSA $A_{LL}$. $A_{LL}$ and its uncertainty were calculated
from the results of the DSSV~2008 global fit~\citep{deFlorian:2008mr}
combined with $P_{h\perp}$ dependence from a fit to recent proton
data~\citep{Avakian:2010ae}. The $A_{LL}$ uncertainty also includes
a contribution from the longitudinal virtual-photon cross section,
which was calculated using the SLAC-R1999 parametrization~\citep{Abe:1998ym}.
The $A_{LT}$ results for $^{3}$He and the $A_{LL}$ correction applied
to the data are shown in Fig.~\ref{fig:He--azimuthal}. Combining
the data from all four $x$\nobreakdash-bins, we have observed a
positive asymmetry with 2.8\textbf{~$\sigma$} significance for $\pi^{-}$
production on $^{3}$He, while the $\pi^{+}$ asymmetries are consistent
with zero. 

The systematic uncertainties in our measurements due to acceptance,
detector response drift and target density fluctuations were suppressed
to a negligible level by the fast beam\nobreakdash-helicity reversal.
With the addition of the frequent target spin reversal, the contributions
from the beam\nobreakdash-SSA $A_{LU}$ and the target\nobreakdash-SSA
$A_{UT}$ were canceled in the extraction of $A_{LT}^{\cos\left(\phi_{h}-\phi_{S}\right)}$.
The dominant systematic effect for the lower $x$\nobreakdash-bins
was the contamination from photon induced charge\nobreakdash-symmetric
$e^{\pm}$ pair production, in which the $e^{-}$ was detected in
the BigBite spectrometer. The yield of $(e^{+},\,\pi^{\pm})$ coincidences
was measured by reversing the magnetic field of the BigBite spectrometer~\citep{Qian:2011py}.
Since the measured asymmetry of the background was consistent with
zero, the contamination was treated as a dilution. Bin centering ($\left|\delta A_{LT}/A_{LT}\right|\leq14\%$)
and radiative ($\left|\delta A_{LT}\right|\leq0.1\%$) effects were
estimated with an adapted {\sc SIMC} Monte Carlo simulation~\citep{Asaturyan:2011mq}
and {\sc polrad2}~\citep{Akushevich:1997di}. Other noticeable systematic
uncertainties include the $\pi^{-}$ contamination in the electron
sample from the BigBite spectrometer ($\left|\delta A_{LT}\right|\leq0.1\%$),
the kaon contamination in the pion sample from the HRS ($\left|\delta A_{LT}\right|\leq0.1\%$),
and the beam and target polarimetry ($\left|\delta A_{LT}/A_{LT}\right|\leq5\%$,
each). Finally, uncertainties in the Cahn ($A_{UU}^{\cos\phi_{h}}$)
and Boer\nobreakdash-Mulders ($A_{UU}^{\cos2\phi_{h}}$) effects
on the unpolarized cross section~\citep{Barone:2010zz} induce relative
systematic uncertainties $\left|\delta A_{LT}/A_{LT}\right|\le10\%$
and $5\%$, respectively. The contamination in identified SIDIS events
from decays of diffractively produced $\rho$ mesons, estimated to
range from 3\%\nobreakdash-5\% (5\%\nobreakdash-10\%) for $\pi^{+}$
($\pi^{-}$) by {\sc pythia6.4}~\citep{Sjostrand:2006za}, was not
corrected, consistent with previous experimental analyses~\citep{:2009ti,Airapetian:2010ds,Avakian:2010ae,Qian:2011py}.
Experimental information regarding the subleading-twist $\cos\phi_{S}$
and $\cos\left(2\phi_{h}-\phi_{S}\right)$ moments of $A_{LT}$ is
rather scarce. However, existing evidence for the suppression of subleading-twist
effects in other observables of inclusive and semi-inclusive DIS in
the kinematic region of this experiment~\citep{Leader:2006xc,Blumlein:2010rn,Asaturyan:2011mq}
supports the leading-twist interpretation presented in this Letter.
Therefore, the potential systematic effect of these terms on the extraction
of the leading-twist $\cos(\phi_{h}-\phi_{S})$ moment is expected
to be small compared to the statistical uncertainties of the present
data, and is not included in the quoted systematic uncertainty. Future
high-precision SIDIS data covering a broader $Q^{2}$ range will enable
an accurate determination of the subleading-twist $A_{LT}$ moments~\citep{Gao:2010av,Anselmino:2011ay}.

The neutron asymmetry was extracted from the $^{3}$He asymmetry using
the effective polarization approximation, given by 
\begin{equation}
A_{LT}^{n}=\frac{1}{\left(1-f_{p}\right)P_{n}}\left(A_{LT}^{^{3}\textrm{He}}-f_{p}A_{LT}^{p}P_{p}\right),
\end{equation}
where the proton dilution factor $f_{p}\equiv2\sigma_{p}/\sigma_{^{3}\text{He}}$
was measured with unpolarized $^{3}$He and hydrogen gas targets in
identical kinematics, including the uncertainties from spin\nobreakdash-independent
final\nobreakdash-state interactions (FSI)~\citep{Qian:2011py}.
The effective neutron and proton polarizations in $^{3}$He are given
by $P_{n}=0.86_{-0.02}^{+0.036}$ and $P_{p}=-0.028_{-0.004}^{+0.009}$~\citep{Zheng:2003un},
respectively. Because of the small proton polarization and a scarcity
of existing data, no $A_{LT}^{p}$ correction was applied to our results.
The allowed range of $A_{LT}^{p}$ was estimated from COMPASS data~\citep{Parsamyan:2010se},
which resulted in a systematic uncertainty in $A_{LT}^{n}$ of less
than $30\%$ of the statistical uncertainty. Target single\nobreakdash-spin-dependent
FSI effects on the DSA were canceled by the frequent target spin flips,
resulting in negligible uncertainty in the extracted $A_{LT}$.

\begin{figure}
\includegraphics[bb=0in 0in 9.4000000000000004in 5.3499999999999996in,clip,width=8.8cm]{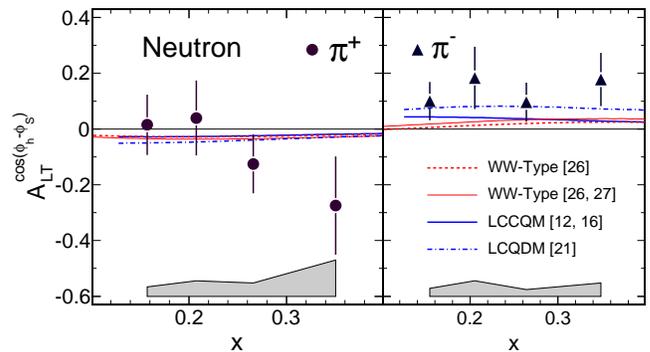}\caption{\textbf{\label{fig:Neutron ALT}}Neutron $A_{LT}^{\cos\left(\phi_{h}-\phi_{S}\right)}$
azimuthal asymmetry for positive (left) and negative (right) charged
pions vs $x$. }
\end{figure}
The results are shown in Fig.~\ref{fig:Neutron ALT} and are compared
to several model calculations, including WW\nobreakdash-type approximations
with parametrizations from Ref.~\citep{Kotzinian:2006dw} and Ref.~\citep{Kotzinian:2006dw,Prokudin.DraftPaper},
a light\nobreakdash-cone constituent quark model (LCCQM)~\citep{Pasquini:2008ax,Boffi:2009sh}
and a light\nobreakdash-cone quark\nobreakdash-diquark model~(LCQDM)~evaluated
using approach two in Ref.~\citep{Zhu:2011zz}. While the extracted
$A_{LT}^{n}\left(\pi^{+}\right)$ is consistent with zero within the
uncertainties, $A_{LT}^{n}\left(\pi^{-}\right)$ is consistent in
sign with these model predictions but favors a larger magnitude. Sizable
asymmetries could be expected for future experiments, including corresponding
SIDIS asymmetries on a proton target and the double\nobreakdash-polarized
asymmetry in Drell\nobreakdash-Yan dilepton production. While the
$\pi^{+}$ and $\pi^{-}$ data are consistent with the interplay between
S\nobreakdash-P and P\nobreakdash-D wave interference terms predicted
by the LCCQM and LCQDM models, the magnitude of the measured $\pi^{-}$
asymmetry suggests a larger total contribution from such terms than
that found in the LCCQM. The larger magnitude of the data compared
to the WW\nobreakdash-type calculations suggests either a different
$P_{h\perp}$ dependence of $A_{LT}$ than assumed in the calculations,
a significant role for subleading\nobreakdash-twist effects, or both.
The statistical precision and kinematic coverage of the present data
cannot distinguish between these scenarios. It is worth noting that
the sign of $A_{LT}^{n}\left(\pi^{-}\right)$ is opposite to the sign
of the $A_{UL}^{\sin2\phi_{h}}$ asymmetry in $\pi^{+}$ production
on the proton measured by the CLAS collaboration~\citep{Avakian:2010ae}.
This observation is consistent with many models which support that
$g_{1T}^{u}$ and $h_{1L}^{\perp u}$ have opposite signs~\citep{Lorce:2011zt}.

In conclusion, we have reported the first measurement of the DSA $A_{LT}^{\cos\left(\phi_{h}-\phi_{S}\right)}$
in SIDIS using a polarized electron beam on a transversely polarized
$^{3}$He target. The neutron $A_{LT}$ was also extracted for the
first time using the effective polarization approximation. Systematic
uncertainties were minimized by forming the raw asymmetry between
beam\nobreakdash-helicity states with minimal charge asymmetry due
to the fast helicity reversal. A positive asymmetry was observed
for $^{3}\text{He}\left(e,e'\pi^{-}\right)X$ and $n\left(e,e'\pi^{-}\right)X$,
providing the first experimental indication of a nonzero $A_{LT}$,
which at leading twist leads to a nonzero $g_{1T}^{q}$. When combined
with measurements on proton and deuteron targets, these new data will
aid the flavor-decomposition of the $g_{1T}^{q}$ TMD PDFs. This work
has laid the foundation for the future high\nobreakdash-precision
mapping of $A_{LT}$ following the JLab 12 GeV upgrade~\citep{Gao:2010av}
and at an electron-ion collider~\citep{Anselmino:2011ay}, which
will provide a comprehensive understanding of the $g_{1T}^{q}$ TMD
PDF and the subleading-twist effects.
\begin{table*}
\centering

\begin{tabular}{|c|c|c|c|c|c|c|c|c|} 
\hline
 $x$  & $Q^{2}$ & $y$ & $z$ & $P_{h\perp}$ &  $^3$He $A_{LT}^{\cos\left(\phi_{h}-\phi_{S}\right)}$  &  $^3$He $A_{LT}^{\cos\left(\phi_{h}-\phi_{S}\right)}$  &  neutron $A_{LT}^{\cos\left(\phi_{h}-\phi_{S}\right)}$ &  neutron $A_{LT}^{\cos\left(\phi_{h}-\phi_{S}\right)}$  \tabularnewline 
      & GeV$^2$ &     &     & GeV          & $\pi^+$          &  $\pi^-$          &  $\pi^+$          &  $\pi^-$           \tabularnewline 
\hline
$0.156$  &  $1.38$ &  $0.81$ &  $0.50$ &  $0.43$  &  $+0.003 \pm 0.020 \pm 0.004$  &  $+0.030 \pm 0.020 \pm 0.008$  &  $+0.02 \pm 0.11 \pm 0.03$  &  $+0.10 \pm 0.07 \pm 0.03$ \tabularnewline
$0.206$  &  $1.76$ &  $0.78$ &  $0.52$ &  $0.38$  &  $+0.005 \pm 0.016 \pm 0.004$  &  $+0.032 \pm 0.019 \pm 0.007$  &  $+0.04 \pm 0.13 \pm 0.06$  &  $+0.18 \pm 0.11 \pm 0.06$ \tabularnewline
$0.265$  &  $2.16$ &  $0.75$ &  $0.54$ &  $0.32$  &  $-0.019 \pm 0.015 \pm 0.004$  &  $+0.024 \pm 0.017 \pm 0.004$  &  $-0.13 \pm 0.11 \pm 0.05$  &  $+0.10 \pm 0.07 \pm 0.02$ \tabularnewline
$0.349$  &  $2.68$ &  $0.70$ &  $0.58$ &  $0.24$  &  $-0.025 \pm 0.015 \pm 0.005$  &  $+0.034 \pm 0.017 \pm 0.006$  &  $-0.27 \pm 0.18 \pm 0.13$  &  $+0.18 \pm 0.10 \pm 0.05$ \tabularnewline
\hline
\end{tabular}

 \caption{Central kinematics\textcolor{black}{{} and }asymmetr\textcolor{black}{y
results.} The format for $^{3}$He and neutron $A_{LT}^{\cos\left(\phi_{h}-\phi_{S}\right)}$
asymmetries follows {}``central value'' $\pm$ {}``statistical
uncertainty'' $\pm$ {}``systematic uncertainty'' .}
\end{table*}

\begin{acknowledgments}
We acknowledge the outstanding support of the JLab Hall A technical
staff and Accelerator Division in accomplishing this experiment. This
work was supported in part by the U. S. National Science Foundation,
and by U.S. DOE contract DE\nobreakdash-AC05\nobreakdash-06OR23177,
under which Jefferson Science Associates, LLC operates the Thomas
Jefferson National Accelerator Facility. 
\end{acknowledgments}
\bibliographystyle{apsrev4-1-mod}
\bibliography{E06010ALT}

\end{document}